\documentstyle[psfig]{mn2e}

\def\ie{i.e$.$~} \def\eg{e.g$.$~} \def\etal{et al$.$~} \def\cm3{\rm cm^{-3}}
\def\Msunyear{M_\odot\,{\rm yr^{-1}}} \def\kms{\rm km\,s^{-1}}
\def\simg{\mathrel{%
      \rlap{\raise 0.511ex \hbox{$>$}}{\lower 0.511ex \hbox{$\sim$}}}}
\def\siml{\mathrel{%
      \rlap{\raise 0.511ex \hbox{$<$}}{\lower 0.511ex \hbox{$\sim$}}}}
\def\epsi{\varepsilon_i} \def\epsB{\varepsilon_B} \def\epse{\epsilon_e}
\def\Mesz{M\'esz\'aros~}

\begin{document}

\title[GRB Afterglows 990123 and 021211]
       {Analysis of Two Scenarios for the Early Optical Emission of the GRB Afterglows 990123 and 021211}

\author[A. Panaitescu \& P. Kumar]{A. Panaitescu and P. Kumar \\
       Department of Astronomy, University of Texas, Austin, TX 78712}

\maketitle

\begin{abstract}

 The optical light-curves of GRB afterglows 990123 and 021211 exhibit a steep decay at 100--600
seconds after the burst, the decay becoming slower after about 10 minutes. We investigate two 
scenarios for the fast decaying early optical emission of these GRB afterglows. In the 
{\sl reverse-forward shock} scenario, this emission arises in the reverse shock crossing the GRB 
ejecta, the mitigation of the light-curve decay occurring when the forward shock emission overtakes 
that from the reverse shock. Both a homogeneous and wind-like circumburst medium are considered. 
In the {\sl wind-bubble} scenario, the steeply decaying, early optical emission arises from the 
forward shock interacting with a $r^{-2}$ bubble, with a negligible contribution from the reverse 
shock, the slower decay starting when the blast wave reaches the bubble termination shock and 
enters a homogeneous region of the circumburst medium. 

 We determine the shock microphysical parameters, ejecta kinetic energy, and circumburst density 
which accommodate the radio and optical measurements of the GRB afterglows 990123 and 021211. 
We find that, for a homogeneous medium, the radio and optical emissions of the afterglow 990123 can 
be accommodated by the reverse-forward shock scenario if the microphysical parameters behind the 
two shocks differ substantially. A wind-like circumburst medium also allows the reverse-forward 
shocks scenario to account for the radio and optical properties of the afterglows 990123 and 021211, 
but the required wind densities are at least 10 times smaller than those of Galactic Wolf-Rayet 
stars. The wind-bubble scenario requires a variation of the microphysical parameters when the 
afterglow fireball reaches the wind termination shock, which seems a contrived feature.

\end{abstract}

\section{Introduction}

 There are currently two GRB afterglows for which a fast falling-off optical emission was detected
at early times, only $\sim 100$ seconds after the burst. The general consensus is that this emission 
arises from the GRB ejecta which is energized by the reverse shock ({\bf RS}) crossing the ejecta and 
caused by the interaction of the ejecta with the circumburst medium ({\bf CBM}). This interaction also 
drives a forward shock ({\bf FS}) energizing the swept-up CBM, to which the later afterglow emission 
is attributed (the "reverse-forward shock" scenario).
  
 The RS emission was first calculated by \Mesz \& Rees (1997), who considered the cases of a 
frozen-in and turbulent magnetic field in the ejecta, and showed that, in either case, a bright 
optical emission ($m_V \sim 9$) is obtained at the end of the burst. 
\Mesz \& Rees (1999) extended their previous calculations of the RS emission to a radiative evolution
of the fireball Lorentz factor and pointed out the importance of spectral information in constraining 
the RS dynamics and the magnetic field origin from the observed $t^{-2}$ power-law decay of the very 
early optical light-curve of the afterglow 990123 (Akerlof \etal 1999). They also pointed out the 
possibility that optical flashes arise in the same internal shocks which generate the burst emission.

  Sari \& Piran (1999) have shown that, if the peak frequency of the RS emission is assumed to be
in the optical at the time when the optical emission of the afterglow 990123 peaks (50 seconds 
after the burst), then the expected softening of the RS emission and self-absorption effects can 
account for the radio flare reported by Kulkarni \etal (1999a). Kobayashi \& Sari (2000) confirm 
the RS interpretation of this radio flare through numerical calculations of the RS dynamics.

 Chevalier \& Li (2000) have presented calculations of the RS synchrotron emission until it crosses 
the GRB ejecta, for the case of a wind-like CBM. For their choice of a high magnetic field parameter, 
the RS cooling frequency falls well below the optical domain, which leads to a RS optical emission 
much dimmer than that observed for the afterglow 990123 at its peak (40 seconds after the burst).
Furthermore, such a low cooling frequency implies that the early afterglow optical emission should
cease when the RS has crossed the ejecta shell, \ie at the peak time of the RS emission. Since this
is in contradiction with the observations of the afterglow 990123, Chevalier \& Li (2000) have 
concluded that a wind-like CBM cannot explain the early optical emission of the afterglow 990123.

 Constraints on the fireball initial Lorentz factor have been obtained by Soderberg \& Ramirez-Ruiz
(2003) for several afterglows by comparing the observed radio emission at $\sim 1$ day with the 
model RS emission, under the assumption that the RS magnetic field and typical electron energy 
parameters (which we shall call {\bf microphysical parameters}) are those determined for the FS from 
fits to the broadband emission of those afterglows (Panaitescu \& Kumar 2001). Zhang, Kobayashi \& 
\Mesz (2003) have noted that the ratios of the RS and FS peak fluxes, and peak and cooling frequencies 
depend only on the fireball initial Lorentz factor and the ratio of the magnetic fields, to develop 
a method of constraining these two quantities, and have shown that the optical emission of the afterglow 
990123 requires a magnetized outflow. 

 In this work we use the general properties (flux, epochs during which power-law decays are observed,
decay slopes -- see Table 1) of the radio and optical emissions of the afterglows 990123 and 021211 
to constrain the ejecta (isotropic-equivalent) kinetic energy, CBM density, and the microphysical 
parameters for the reverse-forward shock scenario (\S\ref{RF}), for either a homogeneous or wind-like 
profile of the CBM. In contrast with other works, we take into account all constraints arising from 
the radio and optical measurements of the GRB afterglows 990123 and 021211 and we do not assume certain 
values for any of the model parameters.
 We also investigate a "wind-bubble" scenario (\S\ref{BB}), where all the radio and optical emission 
arises in the FS, with a negligible contribution from the RS (which is verified numerically), the 
mitigation of the optical decays observed in the afterglows 990123 and 021211 being due to the FS 
crossing the bubble termination shock, i.e. transiting from a wind-like CBM structure to a homogeneous 
region of shocked wind. 
 For both scenarios, we consider either adiabatic or radiative dynamics, the resulting microphysical 
parameters being checked for consistency with the assumed dynamical regime.

\section{Physical Parameters at the Ejecta Shock-Crossing Radius}
\label{dynamics}

 We begin by calculating the spectral properties (break frequencies and peak flux) of the RS emission 
at the radius $R_+$ where the RS finishes crossing the ejecta shell and the injection of fresh electrons 
by the RS ceases. Because most of the ejecta kinetic energy has been transferred to the forward shock
at $R_+$, the radius $R_+$ marks the onset of a steeper power-law decrease of the bulk Lorentz factor 
$\Gamma$ with radius. After $R_+$, the spectral properties of the RS emission can be calculated from the 
adiabatic evolution of the electrons and magnetic field. The spectral properties of the FS emission can 
also be calculated from those at $R_+$ or directly from the dynamics of the fireball after $R_+$ 
(\ie without passing through the parameters at $R_+$) if the shock dynamics is adiabatic. 

 Each shock compresses the fluid ahead of it by a factor $4 \Gamma' + 3$, where $\Gamma'$ is the Lorentz 
factor of the shocked fluid as measured in the frame of the yet unshocked gas, and heats it to a energy 
per particle equal to $\Gamma'-1$. Therefore, the pressure equality at the contact discontinuity which 
separates the shocked ejecta and CBM, implies that
\begin{equation}
 (4 \Gamma' + 3) (\Gamma' - 1) n_{ej} = (4 \Gamma + 3)  (\Gamma - 1) n 
\label{G1}
\end{equation}
where $\Gamma$ is the Lorentz factor of the shocked fluid in the laboratory frame, and $n_{ej}$, $n$ 
are the proton number densities of the unshocked ejecta and of the CBM, respectively, each measured 
in the corresponding comoving frame. From addition of velocities in special relativity,
\begin{equation}
 \Gamma' = \Gamma_0 \Gamma (1 - \beta_0 \beta) \simeq 
          \frac{1}{2} \left( \frac{\Gamma}{\Gamma_0} + \frac{\Gamma_0}{\Gamma} \right) 
\label{G2}
\end{equation}
where $\Gamma_0$ is the initial Lorentz factor of the ejecta, $\beta$ denotes velocities and 
$\Gamma_0 \gg 1$, $\Gamma \gg 1$ were used in the approximation. 
Substituting equation (\ref{G2}) in (\ref{G1}), one obtains a fourth-degree equation for $\Gamma$,
which can be cast in the form
\begin{equation}
 \frac{(\Gamma_0 - \Gamma)^2}{\Gamma \Gamma_0} \left[  \frac{(\Gamma_0 + \Gamma)^2}{\Gamma \Gamma_0} 
               - \frac{1}{2} \right] n_{ej} = 4 \Gamma^2 n \;.
\label{G3}
\end{equation} 
Because $\Gamma \leq \Gamma_0$, the first term in the square bracket is at least eight times larger
than the last term. Ignoring the $1/2$ term, the solution of equation (\ref{G3}) is 
\begin{equation}
 \Gamma = \frac{\Gamma_0}{\left[ 1 + 2 \Gamma_0 \left( \frac{\displaystyle n}
                   {\displaystyle n_{ej}} \right)^{1/2} \right]^{1/2}} \;.
\label{G4}
\end{equation}
The limiting cases for the Lorentz factor are
\begin{equation}
 \Gamma = \left\{ \begin{array}{ll} 
                  \Gamma_0 & \frac{\displaystyle n_{ej}}{\displaystyle n} \gg 4 \Gamma_0^2 \\
                  \left( \frac{\displaystyle \Gamma_0}{\displaystyle 2} \right)^{1/2} 
                      \left( \frac{\displaystyle n_{ej}}{\displaystyle n} \right)^{1/4} & 
                      \frac{\displaystyle n_{ej}}{\displaystyle n} \ll 4 \Gamma_0^2 
               \end{array} \right.  \;.
\label{G5}
\end{equation}
Therefore, in the limit of very dense ejecta, the Lorentz factor of the shocked fluid is the same as
that of the unshocked ejecta, while for more tenuous ejecta, $\Gamma$ depends on the ratio of the
comoving densities. 
 Note that the ratio $n_{ej}$ and $n_{cmb}$ changes with the fireball radius $R$:
\begin{equation}
  n_{ej} = \frac{E}{4 \pi m_p c^2 \Gamma_0 (\Gamma_0 \Delta) R^2} \;,\; n = A R^{-s} 
\end{equation}
where $E$ is the fireball ejecta energy (or, if the outflow is collimated, its isotropic equivalent),
$m_p$ is the proton's mass, $\Delta \ll R$ is the ejecta geometrical thickness measured in the laboratory 
frame (thus $\Gamma_0 \Delta$ is the comoving frame thickness), and we restricted our calculations to two 
simple radial structures of the CBM, either homogeneous ($s=0$) of particle density $n$ or a $R^{-2}$ 
stratification ($s=2$) corresponding to a wind expelled by the GRB progenitor. In the latter case, 
$A = 3 \times 10^{35} A_* \; {\rm cm^{-1}}$, where $A_*$ is the mass-loss rate to wind speed ratio, 
normalized to $10^{-5} \Msunyear/ 10^3 \kms$. By denoting 
\begin{equation}
  X = \frac{E}{4 \pi A m_p c^2}
\end{equation}
the ratio of the comoving densities is
\begin{equation}
 \frac{n_{ej}}{n} = \frac{X}{\Delta \Gamma_0^2 R^{2-s}} \;.
\label{n}
\end{equation}
The ejecta-shell thickness, $\Delta$, is the largest of its initial thickness $c \tau$, where $\tau$
is the laboratory frame duration of the GRB ejecta release by their source, and the expansion due to 
a spread in the radial outflow velocity of the ejecta particles. This velocity spread can be either a 
relic of the initial, super-Eddington radiation pressure in the fireball, or to an imperfect collimation
in the radial direction of the ejecta particles at the end of the fireball acceleration. The former
leads to a comoving-frame expansion of the shell at the sound speed, the latter is expected to produce
a spread of order $1/\Gamma_0$ in the ejecta particles direction of motion (\Mesz, Laguna \& Rees 1993). 
In either case, the resulting contribution to the shell thickness evolution is $R/\Gamma_0^2$, therefore
\begin{equation}
 \Delta = \left\{ \begin{array}{ll} 
             c \tau & R < 2 \Gamma_0^2 c \tau  \\ 
            \frac{\displaystyle R}{\displaystyle 2 \Gamma_0^2} & R >  2 \Gamma_0^2 c \tau 
          \end{array} \right.  \;.
\label{D}
\end{equation}
As shown by Kumar \& Panaitescu (2003), the difference between the laboratory frame speeds of the unshocked 
ejecta and that of the RS is
\begin{equation}
 \beta_0 - \beta_{RS} = \frac{1.4}{\Gamma_0^2} \left( \frac{\Gamma_0^2 n}{n_{ej}} \right)^{1/2} 
\label{bdiff}
\end{equation}
for a wide range of the ratio $\Gamma_0^2 n/n_{ej}$. From here, one can calculate the radius $R_+$ at 
which the RS finishes crossing the ejecta shell:
\begin{equation}
  \Delta (R_+) = \int_0^{R_+} (\beta_0 - \beta_{RS})\; dR \;. 
\label{Rcross}
\end{equation}  
Once $R_+$ is known, equations (\ref{G4}), (\ref{n}) and (\ref{D}) give the Lorentz factor $\Gamma$
at the shock-crossing radius $R_+$ and all the properties of the RS and FS emissions at $R_+$,
which can be then extrapolated at $R > R_+$. We proceed by considering separately ejecta shells 
for which $2 \Gamma_0^2 c \tau < R_+$ and $R_+ < 2 \Gamma_0^2 c \tau$. In the former case, the 
ejecta shell undergoes a significant spreading while the RS propagates into it, while in the latter 
case the spreading is negligible. The usual terminology (Sari \& Piran 1995) is that of "thin ejecta" 
for the former and "thick ejecta" for the latter.

\subsection{Thin Ejecta Shell -- $\Delta = R/(2\Gamma_0^2)$}
\label{thin}

 The substitution of equations (\ref{n}) and (\ref{bdiff}) in (\ref{Rcross}) leads to
\begin{equation}
 (s=0):\; R_+ = \left( \frac{\displaystyle 1.6 X}{\displaystyle \Gamma_0^2} \right)^{1/3} \! \! \! \! = 
             0.94 \times 10^{17} \left( \frac{\displaystyle E_{53}}
             {\displaystyle n_0 \Gamma_{0,2}^2} \right)^{1/3} {\rm cm} 
\label{R1}
\end{equation}
\begin{equation}
 (s=2):\; R_+ = \frac{\displaystyle 0.57 X}{\displaystyle \Gamma_0^2} \! =
                  1.0  \times 10^{15} \frac{\displaystyle E_{53}}
                         {\displaystyle A_* \Gamma_{0,2}^2}\; {\rm cm}
\label{R2}
\end{equation}
where the usual notation $Q_n = Q/10^n$ has been used.
The defining condition for this case, $R_+ > 2 \Gamma_0^2 c \tau$, becomes
\begin{equation}
 \Gamma_0 < \left[ \frac{0.20 X}{(c \tau)^3} \right]^{1/8} = 
              670\; \left( \frac{E_{53}}{n_0 \tau_0^3} \right)^{1/8} \;\; {\rm for} \;\; s=0 
\label{Gmin0}
\end{equation}
\begin{equation}
 \Gamma_0 < \left( \frac{0.29 X}{c \tau} \right)^{1/4} =
              120\; \left( \frac{E_{53}}{A_* \tau_0} \right)^{1/4}  \;\; {\rm for} \;\; s=2  \;.
\label{Gmin2}
\end{equation}
From equation (\ref{n}), the density ratio at $R_+$ is
\begin{equation}
  \frac{n_{ej}}{n} (R_+)= \frac{2X}{R_+^{3-s}} = \left\{ \begin{array}{ll} 
                        1.3\; \Gamma_0^2 & s=0 \\ 3.5\; \Gamma_0^2 & s=2
                              \end{array} \right. \;.
\label{nratio}
\end{equation}
Comparing with equation (\ref{G4}), this shows that at $R_+$ the Lorentz factor of shocked gas is 
in an intermediate regime:
\begin{equation}
 \Gamma_+ \stackrel{s=0}{=} 0.60\, \Gamma_0 \;\;,\;\; \Gamma_+ \stackrel{s=2}{=} 0.70\, \Gamma_0  \;.
\label{G6}
\end{equation}  

 Besides $\Gamma(R_+)$, two other quantities are of interest at the ejecta shock-crossing radius:
the energy of the swept-up CBM, $E_{cbm} (R_+)$, and the corresponding observer frame time, $t_+$. 
The total energy of the shocked CBM is $\Gamma^2$ larger than its rest-mass energy, thus 
\begin{equation}
 E_{cbm} (R) = m_p c^2 \int\limits_0^R 4 \pi r^2  A r^{-s} \Gamma^2(r) \; dr \;.
\label{Ecbm}
\end{equation}
The arrival time $t$ corresponding to the contact discontinuity and the fluid moving toward 
the observer is given by 
\begin{equation}
 t_{CD} (R) = (1+z) \int_0^R \frac{dr}{2c\Gamma^2} \;, 
\label{dt}
\end{equation}
where $z$ is the burst redshift.
Substituting $\Gamma$ from equation (\ref{G4}) in equations (\ref{Ecbm}) and (\ref{dt}), one
obtains 
\begin{equation}
 (s=0):  E_{cbm} (R_+) = 0.25\,E ,\; 
           t_+ = 350\; (1+z) \left( \frac{E_{53}}{n_0 \Gamma_{0,2}^8} \right)^{1/3}  \! \! \! \!{\rm s} 
\label{Et0}
\end{equation}
\begin{equation}
 (s=2):  E_{cbm} (R_+) = 0.34\,E ,\; 
           t_+ = 2.9\; (1+z) \frac{E_{53}}{A_* \Gamma_{0,2}^4}\, {\rm s} \;.
\label{Et1}
\end{equation}
Thus less than half of the initial ejecta energy has been dissipated by the FS by the time when the 
RS crosses the ejecta shell. This means that the shock crossing radius $R_+$ is slightly smaller than 
the usual deceleration radius $R_d$ defined by $E_{cbm}(R_d) = 0.5 E$.

\subsection{Thick Ejecta Shell -- $\Delta = c \tau$}
\label{thick}

 Once again, using equations  (\ref{n}) and (\ref{bdiff}) in (\ref{Rcross}), one obtains that
\begin{equation}
 R_+ \stackrel{s=0}{=} (2.0 X c \tau)^{1/4} = 
                       0.24 \times 10^{17} \left( \frac{\displaystyle E_{53} \tau_0} 
                                        {\displaystyle n_0} \right)^{1/4}\; {\rm cm} 
\label{R3}
\end{equation}
\begin{equation}
 R_+ \stackrel{s=2}{=} (0.51 X c \tau)^{1/2} = 
                       0.52 \times 10^{15} \left( \frac{\displaystyle E_{53} \tau_0} 
                                        {\displaystyle A_*} \right)^{1/2}\; {\rm cm}  \;.
\label{R4}
\end{equation}
The requirement that $R_+ < 2 \Gamma_0^2 c \tau$ leads to the reversed inequalities given in equations
(\ref{Gmin0}) and (\ref{Gmin2}). 
The density ratio (equation \ref{n}) at $R_+$ is
\begin{equation}
  \frac{n_{ej}}{n} (R_+) = \frac{X}{\Gamma_0^2 c \tau R_+^{2-s}} \;.
\label{nr}
\end{equation}
Because in the thick shell case the ejecta density is lower than for a thin shell, the values given
in the $rhs$ of equation (\ref{nratio}) are upper limits for the density ratio. Substituting
equation (\ref{nr}) in the second regime given in equation (\ref{G5}) leads to
\begin{equation}
 (s=0) : \Gamma_+ =  \left[ \frac{X}{32(c \tau)^3} \right]^{1/8} =
                             530\; \left( \frac{E_{53}}{n_0 \tau_0^3} \right)^{1/8} 
\label{G7}
\end{equation}
\begin{equation}
 (s=2) : \Gamma_+ =  \left( \frac{X}{4c \tau} \right)^{1/4} =
                             110\; \left( \frac{E_{53}}{A_* \tau_0} \right)^{1/4} \;.
\label{G8}
\end{equation}

 Note from equations (\ref{G7}) and (\ref{G8}) that, for a thick ejecta shell, the Lorentz factor
$\Gamma$ of the shocked ejecta is independent of that of the unshocked ejecta, $\Gamma_0$. Thus, for
a sufficiently tenuous ejecta, the contrast between $\Gamma_0$ and $\Gamma$ can be sufficiently large
that the RS is relativistic (equation \ref{G2}) in the frame of the incoming ejecta. In contrast, in
the thin ejecta case, equations (\ref{G2}) and (\ref{G6}) lead to $\Gamma' - 1 = 0.13$ for $s=0$ and
$\Gamma' - 1 = 0.064$ for $s=2$, \ie the RS propagating in a thin ejecta shell is trans-relativistic.

 The energy dissipated by the FS at $R=R_+$ and the ejecta shell shock-crossing time are
\begin{equation}
  E_{cbm} (R_+) = 0.36\, E \;, \quad t_+ = 0.71\, (1+z)\, \tau 
\label{Et2}
\end{equation}
for either type of medium.
Note that, as for a thin ejecta shell, the shock-crossing radius $R_+$ is close to the usual 
deceleration radius. 
Furthermore, for a thick ejecta shell, the observer frame shock-crossing time $t_+$ is fairly close 
to the laboratory frame duration of the ejecta release. Given that $t_+$ is roughly the timescale
for dissipating the ejecta kinetic energy and that, most likely, the shock-accelerated electrons cool 
faster than the dynamical timescale, we expect that the duration of an external shock GRB is close 
to $t_+$. The simple temporal structure of the GRBs 990123 and 021211 may suggest that they originate 
in an external shock, nevertheless it is entirely possible that both bursts were produced in internal 
shocks occurring in an outflow with a fluctuating ejection Lorentz factor (Rees \& \Mesz 1994). In this 
case, the internal 
shocks take place at the radius $R_{is} \sim \Gamma_{min}^2 c \tau$, where $\Gamma_{min}$ is the Lorentz 
factor of the slower shells; the shocked fluid moves at $\Gamma \sim (\Gamma_{min} \Gamma_{max})^{1/2}$, 
where $\Gamma_{max}$ is the Lorentz factor of the faster shells; therefore the observed burst duration 
is $t_\gamma = R_{is}/(c \Gamma^2) \sim (\Gamma_{min}/\Gamma_{max}) \tau$. Since a high dissipation 
efficiency requires a large contrast between the Lorentz factors of various ejecta shells ($\Gamma_{min} 
\ll \Gamma_{max}$), the burst duration sets only a lower limit on the duration of the ejecta release,
$\tau$. We shall use this constraint when choosing the shock-crossing time $t_+$ for the GRBs 990123 
and 021211 for the case of a thick ejecta shell.

\section{Synchrotron Emission}
\label{radiation}

 The synchrotron emission from either shock at any observing frequency is determined by the peak 
flux $F_p$, the three break frequencies -- absorption $\nu_a$, injection $\nu_i$, and cooling $\nu_c$ 
-- and the slope of the afterglow spectrum between its break frequencies.

 The peak of the $F_\nu$ synchrotron spectrum, which is at the frequency $\nu_p = \min\{\nu_i,\nu_c\}$,
is given by
\begin{equation}
 F_p = \frac{1+z}{4\pi d_L^2(z)} N_e \Gamma P'_p  
\label{Fp}
\end{equation}
where $z$ is the burst redshift, $d_L$ the luminosity distance (in a $H_0 = 70 \kms/Mpc$, $\Omega_M 
= 0.3$, $\Omega_\Lambda = 0.7$ universe), $N_e$ is the number of radiating electrons, and the factor
$\Gamma$ accounts for the average relativistic boost of the comoving frame synchrotron power $P'_p$
per electron at the peak frequency $\nu_p' = \nu_p/\Gamma$. The Doppler boost $\Gamma$ appears at only 
the first power in equation (\ref{Fp}) because the observer receives emission for an area subtending 
an angle of $\Gamma^{-1}$ radians, \ie from a fraction $\Gamma^{-2}$ of the total number of electrons
$N_e$. Since we will be interested in the early time afterglow emission, the possible collimation of 
the outflow is not an issue here. After the shock crossing radius $R_+$, the number of electrons energized 
by each shock is
\begin{equation}
 N_e^{(RS)} = \frac{E}{m_p c^2 \Gamma_0} \;, \quad  N_e^{(FS)} = \frac{4 \pi}{3-s} A R^{3-s} \;.
\label{Ne}
\end{equation}
The comoving power per electron of equation (\ref{Fp}) is
\begin{equation}
 P'_p = \sqrt{3} \phi_p \frac{e^3}{m_e c^2} B
\end{equation}
where $\phi_p$ is the order-unity coefficient calculated by Wijers \& Galama (1999), $e$ is the
electron charge, and $B$ is the magnetic field strength. The magnetic field is parameterized by 
the fraction $\epsB$ of the post-shock energy density stored in it. Taking into account that
the FS compresses the CBM by a factor $4 \Gamma $ and heats it is to an energy per proton of
$\Gamma m_p c^2$, the magnetic field is
\begin{equation}
 B = \left( 32 \pi\, \epsB\, A R^{-s}\, m_p c^2\, \Gamma^2 \right)^{1/2} \;.
\label{B1}
\end{equation}

 The break frequencies are calculated from the corresponding electron Lorentz factors $\gamma_{a,i,c}$
\begin{equation}
 \nu_{a,i,c} = \frac{3 x_p}{4\pi} \frac{e}{m_e c} \gamma_{a,i,c}^2 B \Gamma
\label{nuaic}
\end{equation}
where $x_p$ is another order-unity factor calculated by Wijers \& Galama (1999) and the last factor
in the $rhs$ is for the average relativistic boost of the fireball emission by its relativistic
expansion. The typical electron Lorentz factor after shock-acceleration is parameterized as
\begin{equation}
 \gamma_i = \epsi \frac{m_p}{m_e} (\Gamma' - 1)
\label{gi}
\end{equation}
where $\Gamma' = \Gamma$ for the FS, while $\Gamma' (R_+)$ for the RS is that given in equation (\ref{G2}) 
for $\Gamma = \Gamma_+$.
We assume that each shock injects in the downstream fluid electrons with a power-law energy
distribution
\begin{equation}
  \frac{dN}{d\gamma} (\gamma > \gamma_i) \propto \gamma^{-p} \;. 
\label{dNdg}
\end{equation}
The acceleration of new electrons by the RS ceases at $R_+$, when all the ejecta have been swept-up, 
but continues at the FS. The cooling electron Lorentz factor is that for which the radiative losses 
timescale is equal to the dynamical timescale:
\begin{equation}
 \gamma_c = 6 \pi \frac{m_e c^2}{\sigma_T} \frac{\Gamma}{(Y + 1) R B^2}
\label{gc}
\end{equation}
where $\sigma_T$ is the Thomson cross-section for electron scattering and $Y$ is the Compton parameter, 
\ie the ratio of inverse Compton to synchrotron power. The Compton parameter is calculated from the 
electron distribution, as described by Panaitescu \& Kumar (2000). The random Lorentz factor of the 
electrons radiating at the self-absorption frequency $\nu_a$ is given by 
\begin{equation}
 \gamma_a = \left\{ \begin{array}{ll} 
            \gamma_p \tau_p^{3/10} & \tau_p < 1 \\ \gamma_p \tau_p^{1/(p+4)} & \tau_p > 1
             \end{array} \right.
           \;,\;\;\; \tau_p = \frac{5 e}{B \gamma_p^5} \frac{N_e}{4\pi R^2}
\label{ga}
\end{equation}
where $\gamma_p = \min\{\gamma_i,\gamma_c\}$ and $N_e$ is given in equation (\ref{Ne}). 

 Equations (\ref{Fp}) -- (\ref{ga}) provide the characteristics of the RS and FS synchrotron emissions 
at $R > R _+$, with the values of $B$, $\gamma_i$ and $\gamma_c$ for the RS calculated at $R = R_+$
from equations (\ref{B1}), (\ref{gi}) and (\ref{gc}), respectively (using equations \ref{R1},
\ref{R2}, \ref{G6}, \ref{R3}, \ref{R4}, \ref{G7}, \ref{G8}) and then extrapolated at $R > R_+$
as described in \S\ref{rs}. Once $\Gamma (R > R_+)$ is known, the fireball radius can be related with 
the observer-frame arrival-time of the photons emitted at that radius and the evolution of the
RS and FS spectral characteristics can be calculated. 

 The slopes of the piece-wise synchrotron spectrum from the broken power-law electron distribution
with energy resulting from shock-acceleration and radiative cooling are described in detail by Sari,
Piran \& Narayan (1998). Note that, for $R > R_+$, the RS electron distribution has a sharp cut-off 
at $\gamma_c$, as the injection of electrons stops when the RS has crossed the ejecta shell, which 
leads to an abrupt switch-off of the RS emission at a given frequency when $\nu_c$ drops below that
frequency.

\subsection{Forward Shock}
\label{fs}

 If the dynamics of the FS is adiabatic, the Lorentz factor $\Gamma_F$ of the FS follows immediately 
from energy conservation $\Gamma_F^2 M_{cbm} c^2 = E$, where $M_{cbm}$ is the mass of the swept-up 
CBM. From here one obtains the Blandford-McKee solution (Blandford \& McKee 1976):
\begin{equation}
 \Gamma_F (R > R_+) = \Gamma_+ \left( \frac{R}{R_+} \right)^{-(3-s)/2} \;.
\label{GFadb}
\end{equation}
Equation (\ref{dt}) also gives the photon arrival time for the emission arising from a patch on the FS
moving at an angle $\Gamma_F^{-1}$ relative to the fireball center -- observer axis, from where most
of the emission arises. Its integration with $\Gamma$ from equation (\ref{GFadb}) leads to 
\begin{equation}
 t_F (R)=\frac{1+z}{4-s} \frac{R_+}{c\Gamma_+^2} \left[\left(\frac{R}{R_+}\right)^{4-s}+1-\frac{s}{2}\right] 
\label{time}
\end{equation} 
where the weak deceleration at $R < R_+$ has been ignored in the calculation of the last term in the $rhs$. 
For $t \gg t_+$, equations (\ref{GFadb}) and (\ref{time}) lead to 
\begin{equation}
 \Gamma_F (t) \propto t^{-(3-s)/(8-2s)} \;,\; R (t) \propto t^{1/(4-s)} \;.
\end{equation}
Substituting in equations (\ref{Ne}), (\ref{B1}), (\ref{gi}), (\ref{gc}) and (\ref{ga}), one obtains
the following scalings
\begin{equation}
 F_p \propto t^{-s/(8-2s)} \;,\; \nu_i \propto t^{-3/2} 
\end{equation}
\begin{equation}
 \nu_c \propto t^{-(4-3s)/(8-2s)} \;,\; \nu_a \propto t^{-3s/(20-5s)} 
\end{equation}
where $\nu_a < \nu_i < \nu_c$ was assumed for the last equation. These scalings are used to calculate
the characteristics of the FS synchrotron emission at any $t > t_+$ from those at $t = t_+$.

 If the swept-up CBM radiates half of its internal energy faster than the dynamical timescale, the 
dynamics of the afterglow is described by $\Gamma_F M = const$, which leads to $\Gamma_F \propto 
R^{-(3-s)}$. Therefore the two extreme regimes of the fireball dynamics are
\begin{equation}
  \Gamma_F^{(adb)} \propto t^{-(3-s)/(8-2s)} \;,\; \Gamma_F^{(rad)} \propto t^{-(3-s)/(7-2s)}
\label{GFrad}
\end{equation}
where the former is the adiabatic case and the latter is a highly radiative regime.
To estimate the effect of high radiative losses, we calculate the dependence of the observed flux on 
the fireball Lorentz factor, $F_\nu \propto \Gamma_F^x$ (with $x$ frequency-dependent), and adjust 
the fluxes obtained in the adiabatic case by a factor 
\begin{equation}
 \frac{F_\nu^{(rad)}}{F_\nu^{(adb)}} (t) = \left( \frac{\Gamma_F^{(rad)}}{\Gamma_F^{(adb)}} \right)^x = 
              \left( \frac{t}{t_+} \right)^{-(3-s)x/[(8-2s)(7-2s)]} \;.
\label{rad}
\end{equation}
The radiative correction factors at $t \sim 0.1$ day are close to those resulting from the expressions 
of the afterglow flux in the adiabatic case (\eg equations B1-B9 and C1-C9 in Panaitescu \& Kumar 2000)
if the fireball energy is decreased by a factor around 10. This means that a highly radiative regime 
corresponds to a fractional energy loss of about 90\% within the first day after the burst. Note that 
for such high radiative losses to occur in an afterglow, FS electrons should acquire a substantial 
fraction of the energy dissipated by the shock and should radiate it quicker that the dynamical timescale, 
which requires a sufficiently high magnetic field, \ie a sufficiently large parameter $\epsB$ and dense 
CBM.

\subsection{Ejecta Shell}
\label{rs}

 We calculate the dynamics of the shocked ejecta by assuming adiabatic dynamics and that the ejecta is
in equilibrium pressure with the energized CBM, whose radial profile (Lorentz factor, density, pressure) 
is described by the Blandford-McKee solution. For adiabatic dynamics, the pressure in the ejecta shell
evolves as 
\begin{equation}
 p_R \propto (R^2 \Delta')^{-a}
\label{pR}
\end{equation}
where $\Delta'$ is the comoving thickness of the ejecta and $a$ is the adiabatic index. From the 
Blandford-McKee solution, the pressure equilibrium at the contact discontinuity implies that
\begin{equation}
 p_R = p_F (\chi_R) = p_F(\chi=1) \chi_R^{-(17-4s)/(12-3s)} 
\end{equation}
where $p_F (\chi_R)$ is the pressure of the shocked CBM at the coordinate $\chi_R$ of the RS, and 
\begin{equation}
 p_F(\chi=1) \propto \Gamma_F^2 n(R) \propto R^{-3}
\label{pF}
\end{equation}
is the pressure immediately behind the FS, the last relation resulting from the dynamics of the FS
(equation \ref{GFadb}). Equations (\ref{pR})--(\ref{pF}) imply that the Blandford-McKee coordinate 
for the RS satisfies
\begin{equation}
 \chi_R \propto (R^{2a-3} \Delta'^a)^{3(4-s)/(17-4s)} \;.
\label{chiR}
\end{equation}  
Equations (\ref{GFadb}) and (\ref{chiR}) and the Blandford-McKee solution for the post-FS Lorentz factor
\begin{equation}
 \Gamma (\chi) \propto \Gamma_F \chi^{-1/2}
\end{equation}  
lead to that the Lorentz factor at the location of the contact discontinuity, \ie the RS Lorentz
factor, evolves as
\begin{equation}
 \Gamma_R \propto R^{-(3-s)/2} (R^{2a-3} \Delta'^a)^{-1.5(4-s)/(17-4s)} \;.
\label{GR}
\end{equation}  

 The adiabatic index $a$ (equation \ref{GR}) of the ejecta is initially $4/3$ if the RS is relativistic, 
which corresponds to the thick ejecta case discussed in \S\ref{thick}, and lower for a 
mildly relativistic shock (thin ejecta -- \S\ref{thin}), decreasing to $5/3$ due to the 
ejecta cooling. Between these limiting cases, the exponents of $R$ and $\Delta'$ given in equation 
(\ref{GR}) change by 0.24 and 0.14, respectively. Such variations do not lead to significant changes 
in the solutions presented in \S\ref{RF}. For ease of further calculations, we will use 
$a=1.5$ in equation (\ref{GR}). 

 The only uncertainty left in the evolution of $\Gamma_R$ is that of the ejecta shell comoving thickness
$\Delta'$. This uncertainty also affects the adiabatic cooling of the electrons and the evolution of
the magnetic field in the ejecta. The evolution of the electron Lorentz factors $\gamma_i$ and $\gamma_c$ 
at $R > R_+$ is 
\begin{equation}
 \gamma_{i,c} \propto V'^{-(a_e-1)} \propto (R^2 \Delta')^{-1/3} 
\label{gic}
\end{equation}
where $V'$ is the comoving frame ejecta volume and an adiabatic index $a_e = 4/3$ for the relativistic
electrons has been used for the last term. Because the ejecta emission switches off when the decreasing
cooling frequency $\nu_c^{(RS)}$ falls below the observing frequency $\nu$, we will search for afterglow 
parameters for which $\nu_c^{(RS)}(t_{max}) > \nu$, where $t_{max}$ is the latest time when the RS
emission was (or thought to have been) observed. Therefore the electrons radiating at frequency $\nu$
cool mostly adiabatically and the ejecta radiative cooling after $R_+$ can be ignored. 
For the magnetic field, the flux-freezing condition yields 
\begin{equation}
  B_\perp \propto (R\Delta')^{-1} \;,\; B_\parallel \propto R^{-2}
\label{B2}
\end{equation}
where $B_\perp$($B_\parallel$) is the magnetic field perpendicular (parallel) to the radial direction 
of the fireball motion.
 
 To assess the effect of the uncertainty of the behaviour of $\Delta'$ on the ejecta synchrotron emission,
we consider two extreme cases: $\Delta' = const$, as could result from the compression ejecta against
the decelerating contact discontinuity, and $\Delta' = R/\Gamma_R$, corresponding to a comoving-frame
expansion of the ejecta shell at a speed comparable to the speed of light. For the former case, relating 
the ejecta radius with the observer time through $R \propto \Gamma_R^2 t$, leads to $R \propto t^{1/4}$ 
for $s=0$ and $R \propto t^{1/2}$ for $s=2$ (just as for the FS). Substituting in equation (\ref{B2}) 
shows that $B_\perp \propto R^{-1}$ decays slower than $B_\parallel$. Then, equations (\ref{Fp}), 
(\ref{nuaic}), (\ref{ga}), and (\ref{gic}) yield
\begin{equation}
 (s=0) \;\; F_p \propto t^{-0.63} \;,\; \nu_{i,c} \propto t^{-0.96} \;,\; \nu_a \propto t^{-0.61}
\end{equation}
\begin{equation}
 (s=2) \;\; F_p \propto t^{-0.75} \;,\; \nu_{i,c} \propto t^{-1.42} \;,\; \nu_a \propto t^{-0.72} \;.
\end{equation}
For an ejecta shell spreading law $\Delta' = R/\Gamma_R$, one obtains that $R \propto t^{1/9}$ for $s=0$ 
and $R \propto t^{1/5}$ for $s=2$. Equation (\ref{B2}) shows that $B_\parallel \propto R^{-2}$ decays 
slower than $B_\perp$, leading to
\begin{equation}
 (s=0) \;\; F_p \propto t^{-0.67} \;,\; \nu_{i,c} \propto t^{-1.19} \;,\; \nu_a \propto t^{-0.41}
\label{scale0}
\end{equation}
\begin{equation}
 (s=2) \;\; F_p \propto t^{-0.80} \;,\; \nu_{i,c} \propto t^{-1.47} \;,\; \nu_a \propto t^{-0.47} \;.
\label{scale2}
\end{equation}
From the above scalings, it can be seen that the temporal index of the break frequencies changes by 
$\sim 0.2$ ($0.05-0.25$) for $s=0$ ($s=2$), while that of the peak flux by 0.05 for either type of 
medium. The solutions presented in \S\ref{RF} vary little between the above assumed behaviours 
of $\Delta'$.

 Equations (\ref{scale0}) and (\ref{scale2}) are used to calculate the characteristics of the RS synchrotron 
emission at $t > t_+$ from those at $t_+$.  The effect of radiative losses on the RS emission is estimated 
in a similar way as for the FS (equation \ref{rad}), by adjusting the fluxes obtained in the adiabatic case 
by a factor which accounts for the faster deceleration of the FS due to the radiative losses. For the highly 
radiative regime, the evolution of the RS Lorentz factor $\Gamma_R$ is calculated as in the adiabatic case 
(equations \ref{pR} and \ref{GR}) but using the scaling of the FS Lorentz factor $\Gamma_F$ with radius
corresponding to the radiative dynamics case (equation \ref{GFrad})\footnotemark. The absorption of the RS 
radio emission in the FS is also taken into account.
\footnotetext{Because the calculation of $\Gamma_R$ makes use of the Blandford-McKee solution for adiabatic 
  dynamics, our calculation of $\Gamma_R$ for radiative dynamics is only a crude approximation}

\section{Reverse-Forward Shock Scenario}

 The formalism presented in \S\ref{dynamics} and \S\ref{radiation} allows the calculation of the 
RS and FS emission at a given observer time and observing frequency. These emissions depend on the dynamics 
of the FS and ejecta shell, \ie on the fireball kinetic energy $E$ and the particle density of 
the CBM ($n$ for a homogeneous medium or $A_*$ for a wind surrounding a massive star). The ejecta-shell 
shock-crossing time $t_+$ also depends on the fireball initial Lorentz factor $\Gamma_0$, for thin ejecta,
or on the duration $\tau$ of the fireball ejection, for thick ejecta. The initial Lorentz factor also 
determines the number of electrons in the ejecta and, therefore, the RS emission. Finally, the RS and FS 
emissions depend on the two microphysical parameters $\epsi$ and $\epsB$ which quantify the typical 
electron energy and the magnetic field. Thus, the RS emission is determined by five parameters in the 
thin ejecta case, or six in the opposite case, while the FS emission depends on four parameters. Note 
that $E$ and $n$ (or $A_*$) determine the emission of both shocks. 

 In this section we determine in the framework of the reverse-forward shock scenario the values of the 
above parameters allowed by the radio and optical emissions of the GRB afterglows 990123 and 021211, 
the only two afterglows for which an optical emission has been detected at early times, $\sim 100$ seconds 
after the burst. 

 Table 1 lists the properties of the burst, optical, and radio emissions of the two afterglows. 
For the afterglow 021211, the optical emission 
is decaying since the first measurement, at $t_1 = 130$ seconds after the burst (Li \etal 2003). For the 
afterglow 990123, the emission begins to decay at $\sim 45$ seconds (Akerlof \etal 1999), after which the 
burst exhibits some variability. This raises the possibility of some energy injection in the RS after 45 
seconds. For this reason, we choose $t_1 = 73$ seconds as the beginning of the afterglow decay, as after 
this epoch the burst exhibits a weaker, decaying emission. 

 In both cases, the early optical emission fall-offs steeper than at later times. For the afterglow 
021211, the transition between these two regimes has been observed: it occurs at $t_* = 550 - 750$ 
seconds (Li \etal 2003). For GRB 990123, the transition is inferred to occur at $t_* = 400 - 700$ seconds 
(Li \etal 2003), \ie around or after the last early optical measurement at $t_2 = 610$ seconds (Akerlof 
\etal 1999) but prior to the next available measurement at $t_3 \sim 4.0$ hours after the burst (Kulkarni 
\etal 1999b).

\begin{table*}
 \begin{minipage}{180mm}
 \caption{Properties of optical and radio emissions used to constrain the afterglow parameters} 
\begin{tabular}{cccccccccccccccccccccc} \hline 
  GRB & z & $t_\gamma$(s) & $t_1$(s) & $F_1$(mJy) & $\alpha_{12}$ & $t_2$(s) & $t_3$ & $\alpha_{34}$ &
               $t_4$(h) & $F_4$($\mu$Jy) & $\beta$ & $t_{5,6,7}$(d) & $F_{5,6,7}$($\mu$Jy)     \\
        & (1) & (2) & (3) & (4) & (5) & (6) & (7) & (8) & (9) & (10) & (11) & (12) & (13)      \\ \hline
 990123&1.6 & 70-100 & 73 & 400 & 1.80 & 610 & 4.2h & 1.10 & 8.3 & 67 & 0.68-0.82 & 0.2,1.2,4.2 & 130,320,68\\
 021211&1.0 & 2-8 & 130 & 4.1 & 1.56 & 650 & 650s & 0.94 & 2.5 & 25 & 0.55-0.98 & 0.1,0.9,3.9 & 84,44,91  \\
               \hline 
\end{tabular}
 (1): burst redshift,
 (2): range of burst duration in various $X$-ray bands,
 (3): selected epoch for early optical measurement,
 (4): $R$-band flux at $t_1$ (Akerlof \etal 1999, Li \etal 2003),
 (5): $t_1 - t_2$ temporal index (Li \etal 2003) of the early optical light-curve, $F_o \propto t^{-\alpha_{12}}$,
 (6): last available measurement or end of the steeply falling-off optical afterglow,
 (7): beginning of slower decaying optical emission,
 (8): $t_3 - t_4$ optical light-curve index (Kulkarni \etal 1999a; Li \etal 2003), $F_o \propto t^{-\alpha_{34}}$,
 (9): selected epoch after $t_3$,
 (10): $R$-band flux at $t_4$ (Fox \etal 2003),
 (11): slope of optical continuum measured after $t_3$ (Holland \etal 2000; Pandey \etal 2003),
          $F_\nu \propto \nu^{-\beta}$,
 (12): selected epochs of radio measurements,
 (13): $2\sigma$ upper limits on radio fluxes (Kulkarni \etal 1999b; Fox \etal 2003).
\end{minipage}
\end{table*}

\subsection{Constraints on the FS and RS Emissions}
\label{RF}

 The slower decaying optical emission lasting for days is naturally attributed to the FS energizing the CBM. 
Its $t^{-1}$ decay implies that the FS injection frequency is below the optical domain at the first epoch,
$t_3$, when the slower decay begins:
\begin{equation}
 \nu_i^{(FS)} (t_3) < 5 \times 10^{14} \; {\rm Hz} \quad ({\rm constraint \; 1}) \;.
\end{equation}
A second constraint is set by the optical flux normalization; for this we choose an epoch $t_4 > t_3$
when the model flux, $F_o^{(FS)}$, is required to be within a factor 3 of the observed flux, $F_o^{(obs)}$,
making allowance for some uncertainty in our calculations\footnotemark:
\begin{equation}
 \frac{1}{3}\, F_o^{(obs)} (t_4) < F_o^{(FS)} (t_4) < 3\, F_o^{(obs)} (t_4) \quad ({\rm constraint \; 2}) \;.
\end{equation}
\footnotetext{This uncertainty factor determines the width of the region of allowed $\epsi-\epsB$ solutions
 in the lower left corner -- upper right corner direction in the figures, and does not affect significantly 
 the conclusions that will be drawn.}

 With the exception of the flare seen at $t = 1.2$ days in the radio afterglow of GRB 990123, there are 
no other detections in the radio down to 0.1 mJy or less. We use the $2\sigma$ upper limits on the radio 
flux at three epochs ($t_{5,6,7}$) spanning the interval 0.1--4 days, to constraint the radio FS 
emission, $F_r^{(FS)}$:
\begin{equation}
 F_r^{(FS)} (t_{5,6,7}) < F_r^{(2\sigma)} (t_{5,6,7})  \quad ({\rm constraint \; 3}) \;.
\end{equation}

 That the early afterglow exhibits a steep fall-off requires that the RS has crossed the ejecta. 
The burst duration, $t_\gamma$, sets a lower bound on the ejecta crossing-time $t_+$, as discussed in 
\S\ref{thick}, thus $t_+$ is constrained by
\begin{equation}
 t_\gamma <  t_+ < t_1  \quad ({\rm constraint \; 4}) \;.
\end{equation}
Furthermore, the steep $t^{-1.6}$ decay of the RS emission at $t_1$ requires that the RS injection frequency 
is below the optical domain at that time:
\begin{equation}
 \nu_i^{(RS)} (t_1) < 5 \times 10^{14} \; {\rm Hz} \quad ({\rm constraint \; 5}) \;.
\end{equation}
Matching the observed flux to within a factor of 3,
\begin{equation}
 \frac{1}{3}\, F_o^{(obs)} (t_1) < F_o^{(RS)} (t_1) < 3\, F_o^{(obs)} (t_1) \quad ({\rm constraint \; 6}) \;.
\end{equation}
is another requirement set on the calculated RS emission. The detection of RS emission until epoch $t_2$, 
when the early observations end (GRB 990123) or when the transition to the FS emission is observed (GRB 
021211), implies that the RS cooling frequency remains above the optical domain until at least epoch $t_2$
\begin{equation}
 \nu_c^{(RS)} (t_2) > 5 \times 10^{14} \; {\rm Hz} \quad ({\rm constraint \; 7}) 
\end{equation}
otherwise, the RS optical emission would exhibit a sharp drop when $\nu_c^{(RS)}$ falls below optical.
Finally, the radio upper limits are imposed on the RS emission as well:
\begin{equation}
 F_r^{(RS)} (t_{5,6,7}) < F_r^{(2\sigma)} (t_{5,6,7})  \quad ({\rm constraint \; 8}) \;.
\end{equation}

 We search for afterglow parameters $(\Gamma_0,E;n/A_*;\epsi,\epsB)$ that lead to FS and RS emissions
satisfying the constraints 1--3 and 5--8, respectively. For GRB 990123, constraint 4 requires that 
$t_+ \sim 70$ s, as for this burst $t_\gamma \sim t_1$. For GRB 021211, the same constraint allows 
that $4\,{\rm s} < t_+ < 130\,{\rm s}$. In the thin ejecta case, $t_+$ determines the ejecta initial
Lorentz factor, $\Gamma_0$ (equations \ref{Et0} and \ref{Et1}). For thick ejecta, $t_+$ determines the 
duration of the ejecta release, $\tau$, (equation \ref{Et2}).  Note that in the case of a thick ejecta 
shell, the reversed inequalities given in equations (\ref{Gmin0}) and (\ref{Gmin2}) provide only a lower 
limit on $\Gamma_0$. Nevertheless, $\Gamma_0$ remains a relevant parameter because its sets the number 
of electrons in the ejecta. 

 The index $p$ of the electron energy distribution (equation \ref{dNdg}) is determined from the exponent 
$\alpha$ of the optical power-law decay, $F_\nu \propto t^{-\alpha}$, for each shock. The available 
measurements of the slope $\beta$ of the optical continuum,  $F_\nu \propto \nu^{-\beta}$, at $t \sim 1$ 
day, constrain the index $p$ through that the intrinsic afterglow spectral slope $\beta_o$ cannot be
larger than observed, as intrinsic spectra harder than observed ($\beta_o < \beta$) can be attributed
to dust reddening in the host galaxy.

\subsection{Results}

 The search for afterglow parameters is done by choosing various combinations of parameters $(E,n)$ 
(for $s=0$) or $(E,A_*)$ (for $s=2$) and by identifying the regions in the $(\epsi, \epsB)$ parameter 
space which satisfy the above constraints. Various values of the initial Lorentz factor $\Gamma_0$ 
satisfying constraint 4 are tried, to maximize the allowed $(\epsi, \epsB)$ parameter range. For thin 
ejecta, the $(E,n/A_*;\epsi,\epsB)$ parameter space for the RS is significantly smaller than for thick 
ejecta. For brevity, we present here only solutions for the latter case.

\begin{figure*}
\begin{minipage}{18cm}
\centerline{\psfig{figure=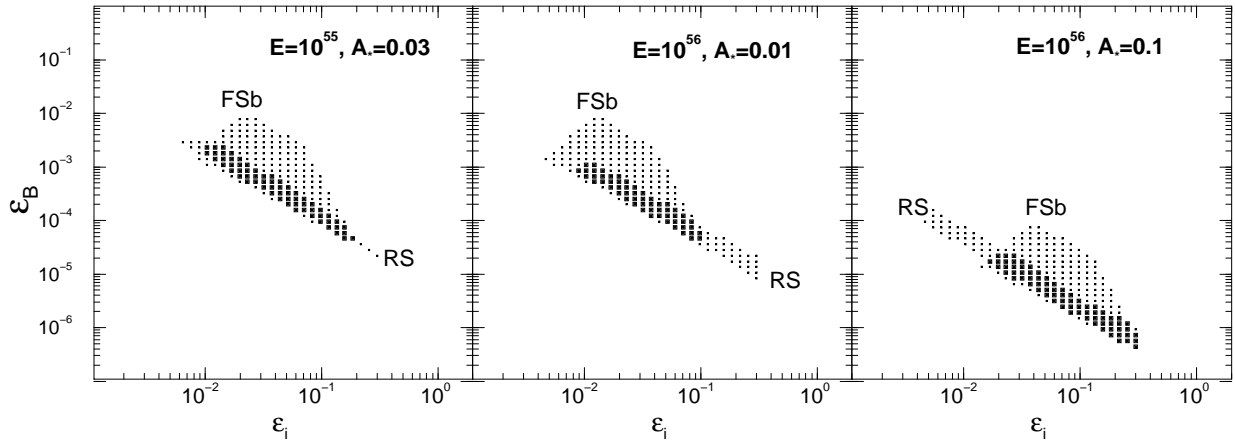,width=16.5cm}}
\caption{{\bf Reverse-forward shock scenario}, {\sl GRB 990123, wind-like CBM}:
  Solutions for radiative dynamics and a CBM shaped by the wind blown by a massive star. 
  The parameter $A_*$ parameterizes the wind particle density $n(r) = \dot{M}/(4\pi m_p r^2 v_w) = 
  3 \times 10^{35} A_* r_{cm}^{-2}$, where $\dot{M}$ is the mass-loss rate and $v_w$ is the wind 
  velocity, with $A_*=1$ corresponding to $\dot{M}/v_w = 10^{-5}\,\Msunyear / 10^3\, \kms$. 
  Higher values of the $A_*$ than shown here, such as those expected for Wolf-Rayet stars 
  ($A_* \simeq 1$), do not allow RS solutions, for either adiabatic or radiative dynamics. 
  Overlapping FS and RS solutions are shown with larger symbols. There are no RS solutions 
  for adiabatic dynamics.  }
\label{jans2}
\end{minipage}
\end{figure*}

\begin{figure*}
\begin{minipage}{18cm}
\centerline{\psfig{figure=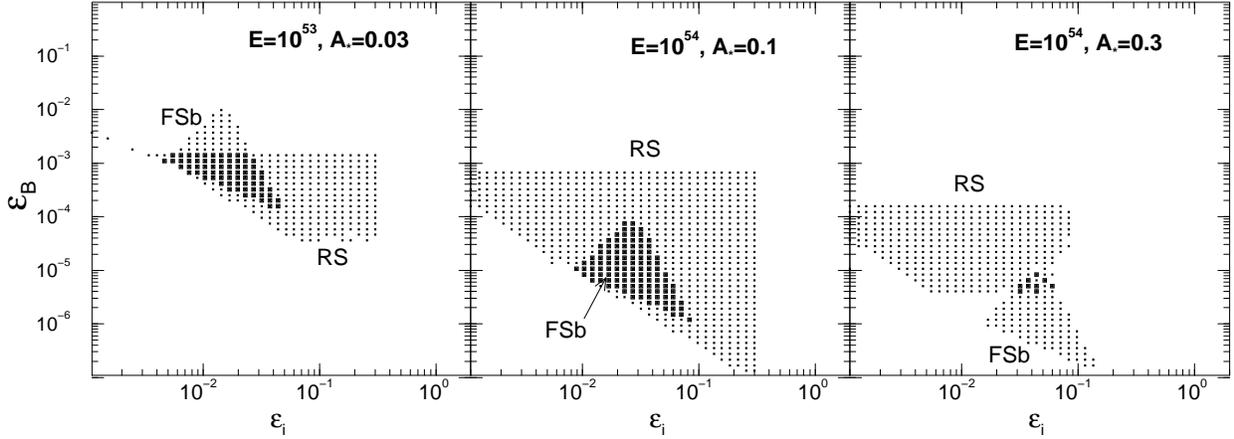,width=16.5cm}}
\caption{{\bf Reverse-forward shock scenario}, {\sl GRB 021211, wind-like CBM}:
  Solutions for radiative dynamics and $t_+ = 100$ s. Adiabatic dynamics also allow solutions with 
  the same microphysical parameters behind the shocks. There are no RS solutions for denser winds 
  (larger $A_*$), or fireball energies $E \siml 3\times 10^{52}$ ergs, for either adiabatic or 
  radiative dynamics. The isotropic-equivalent $\gamma$-ray output of this burst is $E_\gamma = 
  6 \times 10^{51}$ ergs (Vreeswijk \etal 2002).  }
\label{decs2}
\end{minipage}
\end{figure*}

 Figures \ref{jans0}--\ref{decs2} show the RS and FS solutions
$(\epsi,\epsB)$ for various combinations $(E,n)$ or $(E,A_*)$ for which both RS and FS solutions exist. 
For parameters $E$, $n$ (or $A_*$) different by a factor 10 than those shown, the emission from one of 
the shocks, or from both, fails to satisfy the above constraints. For all these three figures, the 
adiabatic index in the shocked ejecta (\S\ref{rs}) was set to $a=1.5$ and the comoving thickness 
of the ejecta was assumed to evolve as $\Delta' = R/\Gamma_R$, where $\Gamma_R$ is the bulk Lorentz 
factor of the shocked ejecta at $R > R_+$. Taking $a = 4/3$, as for a relativistic RS, or assuming
a constant $\Delta'$ lead to a modest change in the solutions shown and leave unaltered the conclusions
below. To assess the effect of radiative losses, the fluxes obtained in the adiabatic case were decreased
as expected from the faster deceleration of $\Gamma$ in the case of radiative dynamics (equation \ref{rad}).

 For the reverse-forward shock scenario we reach the following conclusions about the RS and FS 
parameters: 

\begin{figure*}
\begin{minipage}{18cm}
\centerline{\psfig{figure=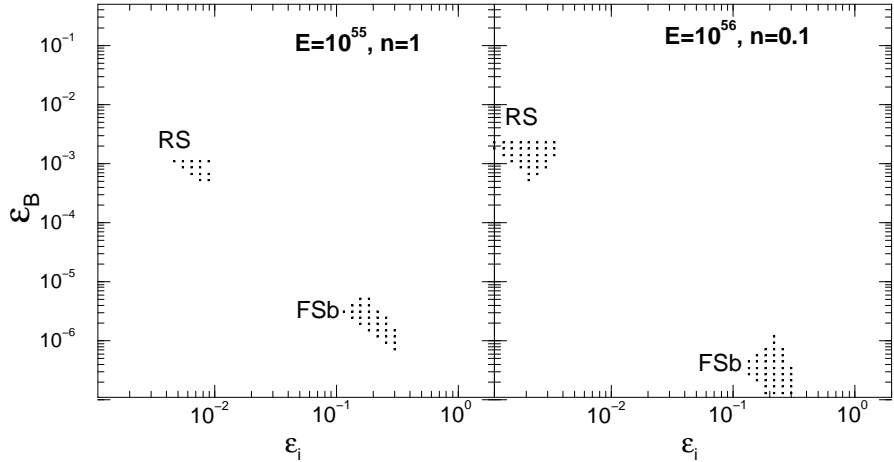,width=12cm}}
\caption{{\bf Reverse-forward shock scenario}, {\sl GRB 990123, homogeneous circumburst medium (CBM)}: 
  Reverse and forward shock microphysical parameters $(\epsi,\epsB)$ for the minimal electron energy 
  and magnetic field energy, satisfying the constraints described in \S\ref{RF}. 
  An uncertainty of a factor 3 is assumed in the calculated optical fluxes, however no uncertainty 
  factor is assigned to the analytical radio fluxes, as they are compared with $2\,\sigma$ observational 
  upper limits.  "RS" and "FSb" denote reverse and forward shock solutions, respectively, the FS solutions 
  corresponding to a cooling frequency above (blueward of) the optical domain.
  The solutions shown are for thick ejecta (\S\ref{thick}), with an observer-frame RS ejecta crossing-time 
  $t_+ = 70$ s, and for highly radiative dynamics (\S\ref{fs}). There are no FS solutions for adiabatic 
  dynamics. Each panel specifies the isotropic-equivalent of the blast-wave kinetic energy, $E$ (in ergs),  
  and CBM particle density, $n$ (in $\cm3$). For values of $(E,n)$ differing by a factor 10 or more than 
  those shown here, the emission of at least one of the shocks becomes incompatible with the observations. 
  For comparison, the isotropic equivalent of the gamma-ray output of this burst is $E_\gamma = 3 \times 
  10^{54}$ ergs (Kulkarni \etal 1999b).  }
\label{jans0}
\end{minipage}
\end{figure*}

(1) {\sl GRB 990123, homogeneous CBM (figure \ref{jans0}):} 
  RS and FS microphysical parameters differ, with $\epsB^{(RS)} \simg 100\, \epsB^{(FS)}$ and 
  $\epsi^{(RS)} < 0.1 \, \epsi^{(FS)}$.
  A RS magnetic field larger than that behind the FS points to an ejecta which was initially magnetized.
  FS solutions correspond to a cooling frequency, $\nu_c^{(FS)}$, above the optical domain. 
  The power-law decay of the FS optical light-curve, $\alpha_{34}$, sets the electron index: 
  $p = (4\alpha_{34} + 3)/3 = 2.47 \pm 0.07$. This implies that the slope of the intrinsic 
  optical continuum is $\beta_o = (p-1)/2 = 0.74 \pm 0.04$, consistent with the value reported 
  by Holland \etal (2000), at $t=1-3$ days after the burst. FS solutions are obtained only 
  for high radiative losses. For the FS parameters shown in figure \ref{jans0}, the ratio of 
  the (inverse-Compton) cooling to injection electron Lorentz factors is $\gamma_c/\gamma_i 
  \sim 10\, (t/0.1\,d)^{2/3}$, therefore the energy given to electrons which cool radiatively 
  ($\gamma > \gamma_c$) is a fraction $f(t) = (\gamma_c/\gamma_i)^{2-p} = 0.34\, (t/0.1\,d)^{-1/3}$ 
  of the total electron energy. If the injected electron distribution extends to arbitrarily 
  high energies, the fraction of the post-shock energy in electrons is $\epse = [(p-1)/(p-2)] 
  \epsi = 3\, \epsi \in (0.3, 1)$. Then, at $t = 0.1$ day, the radiative losses over one dynamical 
  timescale are a fraction $\xi_{rad} (0.1\, d)= f \epse \simeq \epsi \in (0.1,0.3)$ of the FS 
  energy. If the total electron energy does not exceed equipartition ($\epse \leq 0.5$), then
  $\xi_{rad} (0.1\, d)\siml 0.15$. Therefore, for the FS solutions shown in figure \ref{jans0}, 
  the fireball dynamics is between the adiabatic and highly radiative regimes. 

 The ejecta kinetic energy $E \simg 10^{55}$ ergs (if spherical) and the ambient medium density 
$n \simg 0.1\, \cm3$ for which the RS emission accommodates the early optical emission of the afterglow 
990123 are 10 and 50 times larger than their upper limits found by us (Panaitescu \& Kumar 2001) from 
multiwavelength afterglow modelling. The difference is caused by the inclusion in the current 
calculations of the early ($t < t_*$) optical emission, which cannot be accommodated by the RS if 
the fireball energy and CBM density were those determined from modelling the afterglow emission at 
$t > 4$ hours, as the constraints 6 and 8 in \S\ref{RF} cannot be satisfied simultaneously. 
Conversely, for the ($E,n$) values for which the RS emission accounts for the early optical emission, 
the FS solutions identified in this work satisfy the general constraints imposed by observations 
(\S\ref{RF}) but provide a $\chi^2$-wise poorer fit to the afterglow data after 4 hours than 
the best fit obtained numerically with more accurate calculations of radiative losses and the
integration of received emission over the photon equal arrival time surface. 

(2) {\sl GRB 990123, $r^{-2}$ CBM (figure \ref{jans2}):} 
  there are RS and FS solutions with the same microphysical parameters. RS solutions exist only 
  for radiative dynamics. Denser wind environments are found to allow RS solutions if the kinetic 
  energy $E$ is increased, but even for $E = 10^{56}$ ergs, which is 30 times larger than the GRB 
  output, the allowed wind density is well below that of a Wolf-Rayet (WR) star ($A_* \simeq 1$). 
  FS solutions correspond to $\nu_c^{(FS)}$ above the optical domain, therefore the electron 
  index is $p = (4\alpha_{34} + 1)/3 = 1.83 \pm 0.08$, leading to $\beta_o = (p-1)/2 = 0.42 
  \pm 0.04$, \ie significantly harder than any reported measurement. The RS and FS solutions
  shown in figure \ref{jans2} with $\epsi \sim 10^{-2}$ are consistent with high radiative
  losses if electrons reach equipartition ($\epse = 0.5$), because $p < 2$ allows most of the 
  electron energy to be at $\gamma > \gamma_c$. However, for the solutions with $\epsi \sim 0.1$, 
  equipartition energies require a cut-off in the accelerated electron distribution at 
  $\gamma_{cut} < \gamma_c$. The synchrotron characteristic frequency for the $\gamma_{cut}$
  electrons is above the optical domain (thus the cut-off does not affect the optical afterglow), 
  but $\gamma_{cut} < \gamma_c$ implies that all injected electrons are cooling adiabatically, 
  \ie the assumption of high radiative losses becomes invalid for $\epsi \simg 0.1$. 

(3) {\sl GRB 021211, homogeneous CBM :} 
   in our previous study (Kumar \& Panaitescu 2003) we have shown that, in the thin ejecta shell 
   case, the RS magnetic field parameter must be $10^3-10^4$ times larger than for the FS. For a 
   thick ejecta shell, corresponding to a RS shell-crossing time longer than the burst duration 
   (2--8 s) but shorter the time of the first optical measurement (130 s), there are RS and FS 
   solutions with the same microphysical parameters if the dynamics is radiative.

(4) {\sl GRB 021211, $r^{-2}$ CBM (figure \ref{decs2}):} 
  RS and FS may have the same microphysical parameters, for either adiabatic or radiative dynamics. 
  There are also solutions with the same parameter $\epsi$ and a RS magnetic field parameter larger 
  than for the FS, indicative of a frozen-in magnetic field. FS solutions correspond to $\nu_c^{(FS)}$ 
  above the optical domain, leading to $p = 1.59 \pm 0.08$ and $\beta_o = 0.29 \pm 0.04$, which is 
  $2\sigma$ below the hardest slope reported (Pandey \etal 2003, at $t=20$ hours). Because $p < 2$, 
  the dynamics may be radiative for $\epsi < 0.1$ and $\epse = 0.5$, while a significantly lower 
  $\epse$ would ensure an adiabatic dynamical regime. Note that a high $E = 10^{54}$ ergs, more than 
  100 times larger than the burst output, is required by a wind density corresponding to $A_* = 0.3$, 
  \ie slightly below that of a WR star. From the constraints on the FS parameters, Chevalier, Li 
  \& Fransson (2004) have also concluded that a weak wind with $A_* \sim 0.01$ is required for the 
  afterglow 021211, however, in their calculations, the FS cooling frequency was placed below the 
  optical. If we impose the same constraint here, we do not find any solutions for the RS microphysical 
  parameters.

\section{Wind-Bubble Scenario}
\label{BB}

 The similarity of the decay indices of the optical light-curves of the afterglows 990123 and 021211 
before and after $t_* \sim 600$ may suggest that a single mechanism produces the entire optical 
afterglow emission. Time-varying microphysical parameters, including the slope of the power-law 
electron energy distribution, could cause a change in the optical light-curve decay index, however 
such an explanation is in contradiction with the consistency seen in many afterglows between the 
optical spectral slope and light-curve decay index at times of order 1 day.

 If the FS is the only mechanism producing the detected afterglow emission and if microphysical 
parameters are constant, then the non-monotonic behaviour of the optical light-curves of the afterglows 
990123 and 021211 at $t_* \sim 600$ seconds must be tied with the fireball dynamics. The fireball 
dynamics is determined by the ejecta initial energy and CBM density. A substantial energy injection 
can mitigate the afterglow dimming rate, however the energy deposition would have to last for the 
entire duration of the slower power-law decay, \ie until at least a few days, and could lead to a 
too bright radio emission from the RS. 

 Besides energy injection, a sudden variation in the radial profile of the CBM could also alter 
the afterglow behaviour. Such a variation is suggested by the association of GRBs with the death 
of massive stars, which drive powerful winds, and that the modelling of multiwavelength afterglow
measurements starting a few hours after the burst leads to better fits for a homogeneous medium
than a wind. This discrepancy may be resolved if the afterglow does not arise in a freely
expanding wind, but in the environment resulting from the interaction of the wind with the
circumstellar gas (Wijers 2001) or with the winds blown by other stars (Scalo \& Wheeler 2001).
It is then possible that the GRB ejecta run into a CBM whose density profile at
smaller radii is the $r^{-2}$ expected for a uniform, free wind, and closer to uniformity at 
larger distances. Then, if the FS cooling frequency is above the optical domain, the optical 
afterglow light-curve index should decrease by $\delta \alpha = 0.5$ when the wind termination 
shock is reached. The resulting index decrease is slightly smaller than observed, but it is possible 
that deviations from uniformity of the environment outside the unperturbed wind account for the 
difference.

 One important issue for this wind-bubble scenario is under what conditions the wind termination 
shock is located at the radius $R_*$ of the afterglow at the time $t_*$ when the light-curve transits 
from a steeper to a slower decay. From equations (\ref{R2}) and (\ref{time}), one obtains
\begin{equation}
 R_* \sim 1.4\times 10^{16} \; \left( \frac{E_{53}}{A_*} \frac{t_*}{600\;{\rm s}} \frac{2}{1+z} 
                   \right)^{1/2} \; {\rm cm} \;.
\label{Rtrans}
\end{equation} 
$R_*$ is higher by a factor of a few for the wind parameters $A_* \sim 0.1$ which we find for 
this scenario, and by an extra factor  $\simg 10$ for the high ejecta kinetic energy obtained 
for the afterglow 990123. Thus, we shall find that $R_* \sim 0.3$ pc for the afterglow 990123 
and $R_* \siml 0.02$ pc for the afterglow 021211. Because the afterglow radius increases as 
$R \propto t^{1/4}$ at $R > R_*$ and because the slower decay of the afterglows 990123 and 021211 
is seen from $t = t_*$ untill $t \sim$ few days, the uniform part of the CBM must extend up to
at least $5\, R_*$.

 Castor, McCray \& Weaver (1975) have derived the major physical properties of a bubble resulting 
from the interaction of stellar winds with the interstellar gas, taking into account the cooling 
and the diffusion of the interstellar gas into the shocked wind. The radius of the wind termination 
shock $R_t$ can be estimated from the equality of the wind ram pressure and that inside the bubble,
leading to $R_t = 4\, \dot{M}_{-5}^{0.3} v_{w,3}^{0.1} n_0^{-0.3} t_5^{0.4}$ pc, where $\dot{M}_{-5}$ 
is the mass-loss rate in $10^{-5}\, \Msunyear$, $v_{w,3}$ is the wind velocity in km/s, $n_0$ is
the interstellar gas density in $\cm3$, and and $t_5$ is the duration of the wind measured in 
$10^5$ years. From here, the ratio of the contact discontinuity radius $R_{cd}$ to that of the wind 
termination shock is $R_{cd}/R_t = 2.3\, \dot{M}_{-5}^{-0.1} v_{w,3}^{0.3} n_0^{0.1} t_5^{0.2}$. 
Thus for a GRB occurring in a dense cloud ($n > 10^5\, \cm3$), the termination shock radius 
could be at location required by the CBM scenario (equation \ref{Rtrans}) and the shocked wind 
shell could be sufficiently thick. 

 However, WR winds do not interact with the interstellar medium but with the wind expelled during 
the red supergiant (RSG) phase, which collides with the main-sequence phase wind decelerated 
by the interaction with the interstellar medium.
The numerical hydrodynamical calculations of Ramirez-Ruiz \etal (2001) take into account the 
wind history and show that $R_t \sim 0.02$ pc for $n = 1 \,\cm3$ and $t = 10^6$ years. Such a 
termination shock radius is suitable for the wind-bubble scenario and the afterglow 021211,
however the wind-bubble size (0.3 pc) shown by  Ramirez-Ruiz \etal (2001) is surprisingly small,
being 100 times less than that expected from the analytical results of Castor \etal (1975). 
Chevalier \etal (2004) have considered the possibility that a high interstellar pressure may 
stall the bubble expansion. For the external pressure expected in a intense starburst region, 
their numerical simulations lead to a wind shock termination radius $R_t = 0.4$ pc and a contact
discontinuity located at $R_{cd}  = 4\, R_t$, which is about right for the wind-bubble 
scenario and the afterglow 990123.

 Alternatively, the uniformity of the CBM at $R > R_*$ required by the wind-bubble scenario 
might arise from a sudden increase in the wind speed, leading to a inner shock propagating into 
the incoming wind. The self-similar solutions derived by Chevalier \& Imamura (1983) for colliding 
winds show that a thick, uniform density shell forms behind the inner shock if the termination 
shock moves at less than 1\% of the unshocked wind speed, which requires the fluctuation in the 
wind to consist of a decrease in the mass-loss rate by a factor 100 and an increase of the wind 
speed increases by a factor 100 or larger. Such a dramatic change in the wind properties exceeds 
that expected at the transition from a luminous blue variable (LBV) wind ($\dot{M}\simg 10^{-3}\, 
\Msunyear$, $v_w = 200\, \kms$; Garcia-Segura, Mac Low, \& Langer 1996) or a RSG wind ($\dot{M} = 
10^{-4}\, \Msunyear$, $v_w \siml 100\, \kms$; Garcia-Segura, Langer \& Mac Low 1996) to a Wolf-Rayet 
wind ($\dot{M} = 10^{-5}\, \Msunyear$, $v_w \simg 1000\, \kms$). The inner shock speed $v_{sh} = 
v_w^{(WR)}/100 = 10\, \kms$ and the location of the termination shock required by the afterglows 
990123 and 021211 imply a WR lifetime $R_*/v_{sh} = 3\times 10^4$ and $2 \times 10^3$ years, 
respectively, \ie much shorter than predicted by evolutionary models for such stars.

\begin{figure*}
\begin{minipage}{18cm}
\centerline{\psfig{figure=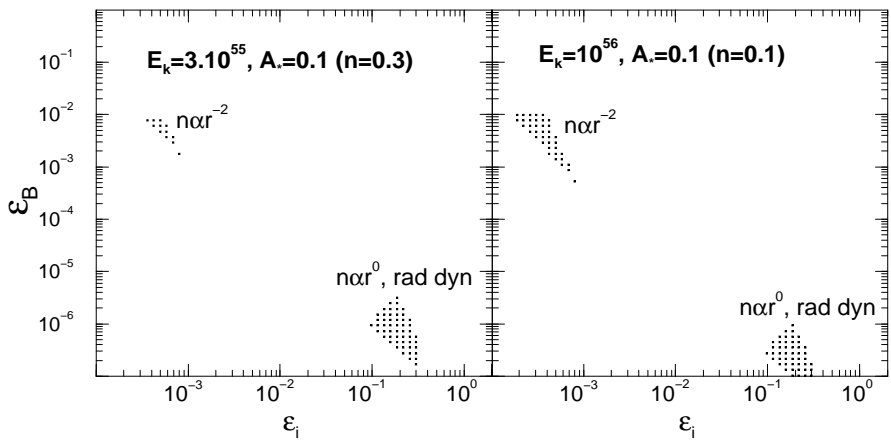,width=12cm}}
\caption{{\bf Wind-bubble scenario}, {\sl GRB 990123}:
  Forward shock microphysical parameters shown separately for the wind-like, inner region of the CBM
  (marked "$n \propto r^{-2}$"), corresponding to the early, fast decaying, optical afterglow ($t < 650$s), 
  and the homogeneous, outer part of the CBM (denoted by "$n \propto r^0$"), where the $t > 4$ hours, 
  slower falling-off emission arises. These solutions were calculated assuming radiative dynamics, 
  which is consistent with the resulting values of the microphysical parameters for $n \propto r^0$. 
  Each panel indicates the fireball kinetic energy $E$, the wind parameter $A_*$, and the resulting
  density $n$ (equation \ref{nout}) of the uniform outer medium. There are no solutions for the 
  the $n \propto r^0$ region for denser winds or lower fireball energies. The lack of overlap between 
  the wind and uniform medium solutions indicate that this scenario requires the parameters for magnetic
  field and minimal electron energy behind the FS to vary when the wind-bubble termination shock is 
  encountered. }
\label{janwb}
\end{minipage}
\end{figure*}

\begin{figure*}
\begin{minipage}{18cm}
\centerline{\psfig{figure=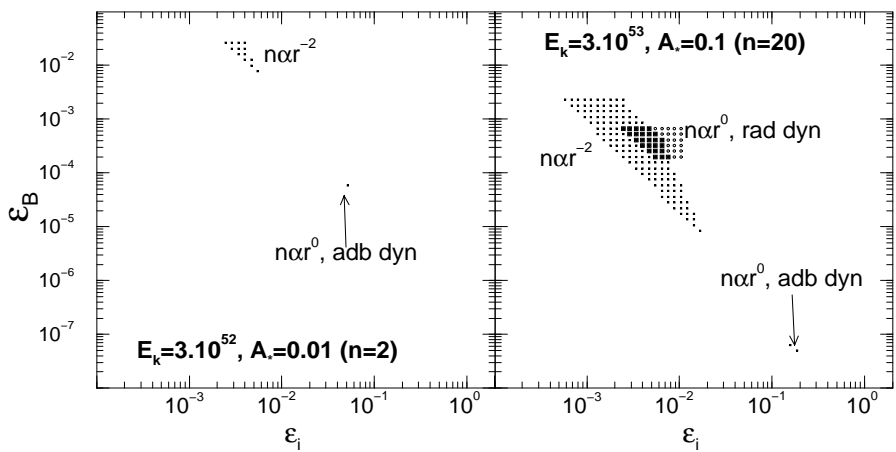,width=12cm}}
\caption{{\bf Wind-bubble scenario}, {\sl GRB 021211}:
 Same as in figure \ref{janwb} but for the afterglow 021211. The allowed regions for the microphysical
 parameters for the wind and uniform portions of the CBM are shown. In the left panel, the assumption 
 of adiabatic dynamics is justified by the resulting total electron energy $\epse = (p-1)\epsi/(p-2) = 
 5 \epsi < 0.1$ for the required electron distribution index $p = 2.25$. In the right panel, for the 
 homogeneous CBM region, neither dynamical regime is consistent with the microphysical parameters shown, 
 thus the correct $n \propto r^0$ solutions should lie somewhere between the two extreme regimes,
 most likely not overlapping with the wind ($n \propto r^{-2}$) solutions.} 
\label{decwb}
\end{minipage}
\end{figure*}

 Having found some support for the wind-bubble scenario in the stalled WR wind-bubble model
of Chevalier \etal (2004), and less so in the interaction between the RSG/LBV and WR winds, we
proceed with testing it against the radio and optical observations of the afterglow 990123 and 
021211. The wind-bubble scenario must satisfy the constraints given in \S\ref{RF},
the first three pertaining to the emission from the fireball interacting with the homogeneous 
portion of the CBM, while the last three referring to the FS propagating in the $r^{-2}$ bubble
(instead of the "RS", as indicated in those equations). In addition, the constraint
\begin{equation}
 \nu_c^{(FS)} (t_1, t_4) > 5 \times 10^{14} \; {\rm Hz}
\label{nuc}
\end{equation}
must be imposed, to explain the decrease of the afterglow dimming rate at $t_*$, when the uniform 
medium is encountered. We note that, for $s=2$, $\nu_c$ increases in time, while for $s=0$ it decreases. 
Thus, if condition (\ref{nuc}) is satisfied, then $\nu_c^{(FS)}$ is above the optical domain for any 
$t \in (t_1,t_4)$.  We also note that this scenario has only four parameters, two ($E$ and $A_*$) for 
the FS dynamics and two ($\epsi$ and $\epsB$) for the emission. The density of the uniform region of 
the CBM is determined by the compression by a factor 4 of the wind density at the location of the 
termination shock. From equation (\ref{Rtrans}), one obtains that this density is
\begin{equation}
  n_{s=0} = \frac{10^{36}\,A_*}{R_{*,{\rm cm}}^2} = 6 \times 10^3\,
          \frac{A_*^2}{E_{53}} \left( \frac{t_*}{600\;{\rm s}} \frac{2}{1+z} \right)^{-1} \; \cm3 \;.
\label{nout}
\end{equation}
The density jump across the termination shock should lead to a brief brightening of the afterglow
(Wijers 2001).
Such a behaviour may have been missed in the afterglow 990123, where there is a gap in the optical 
observations from 10 minutes to 4 hours. It is not seen in the optical measurements of the afterglow
021211 at the "break" time of 10 minutes determined by Li \etal (2003), instead an optical emission 
brighter by 0.5 magnitudes than the double power-law fit used there is seen at 2 hours after the 
burst. The lack of a brightening at the right time in the afterglow 021211 may be problematic for 
the wind-bubble scenario considered here.

 Figures \ref{janwb} and \ref{decwb} display separately the FS microphysical parameters which 
accommodate the observations before and after $t_*$, for a few combinations of fireball energy 
$E$ and wind parameter $A_*$. Smaller values of the former parameter or larger values for the latter 
do not allow the FS to accommodate the $t > t_*$ radio and optical emission of the afterglows
990123 and 021211. Note that, because the RS emission has at least one free parameter ($\tau$ and
$\Gamma_0$), requiring that it does not exceed the measured optical fluxes or the radio upper limits, 
does not constrain the FS parameters. 

 As shown in figure \ref{janwb}, the parameters $\epsi$ and $\epsB$ satisfying the observational 
constraints cannot be constant across $R_*$ for the afterglow 990123, with the electron energy 
parameter increasing by a factor $\simg 100$ and the magnetic field parameter decreasing by a 
factor $\simg 100$ when the fireball crosses the wind termination shock. A similar conclusion is 
reached for the afterglow 021211 (figure \ref{decwb}).

\section{Conclusions}

 We have investigated two scenarios that can account for the behavior of the early optical emission
of GRB afterglows 990123 and 021211, whose light-curves fall-off as $t^{-1.7 \pm 0.1}$ at $t < t_* 
\simeq 600$ seconds, while the decay at $t > t_*$ follows $t^{-1.0 \pm 0.1}$. The first scenario is 
the widely used reverse-forward shock scenario, where the fast decay of the early optical emission 
is attributed to the GRB ejecta energized by the reverse shock (RS), and the slower decaying phase 
is associated with circumburst medium (CBM) swept-up by the forward shock (FS).

 Figure \ref{jans0} shows that, for a homogeneous medium, the reverse-forward shock scenario can 
accommodate the radio and optical measurements of the afterglow 990123 if the ejecta magnetic field 
is $\simg 10$ times larger than in the shocked CBM, which implies an ejecta frozen-in magnetic field 
(\Mesz \& Rees 1997, Zhang \etal 2003), and if the RS parameter for the typical electron energy is 
$\simg 10$ times smaller than for the FS. This differences between the microphysical parameters behind 
the two shocks are too large to be explained away by the inaccuracies in the calculations of the 
afterglow emission presented in \S\ref{radiation}. 

 For a wind-like CBM (figure \ref{jans2}), the reverse-forward shock scenario can explain the major 
properties of the radio and optical emissions of the afterglow 990123 with the same microphysical 
parameters behind both shocks (figure \ref{jans2}), but it requires a wind density corresponding 
to a mass-loss rate to speed ratio less than $10^{-6}\, \Msunyear/ 10^3\, \kms$ (\ie $A_* < 0.1$). 
We obtain a similarly tenuous wind also from modelling the broadband data of the afterglow 990123 
at $t > 0.2$ days, but it should be noted that the best fit with a wind-like CBM provides a poorer 
fit to the data than a uniform medium. 

 Within the framework of the reverse-forward shock scenario, the same microphysical parameters are 
obtained for the afterglow 021211, for a thick ejecta shell and either a uniform or a wind-like CBM 
(the latter case is shown in figure \ref{decs2}). For a uniform CBM, we obtain $n \simg 30\, \cm3$, 
larger than the $n \siml 1\, \cm3$ inferred by us (Kumar \& Panaitescu 2003) for a thin ejecta shell. 
For a wind-like CBM, the afterglow 021211 requires a wind density corresponding to a mass-loss rate 
to speed ratio below $10^{-6}\, \Msunyear/ 10^3\, \kms$, a result similar to that obtained for the 
afterglow 990123.

 The second scenario considered in this work (\S\ref{BB}), that of wind-bubble having a inner $r^{-2}$ 
wind-like region surrounded by a zone of uniform density, is motivated by that the decrease in the dimming 
rate of the optical afterglows 990123 and 021211 seen at $t_* \sim 650$ seconds matches fairly well 
the expectations for such a density profile. The required CBM structure finds support in the scenario 
of WR wind-bubbles stalled by a high interstellar pressure discussed by Chevalier \etal (2004). 
For this scenario to explain the general properties of the radio and optical emissions of the 
afterglows 990123 and 021211, the magnetic field and electron energy parameters would have to 
decrease and increase, respectively, by a factor of about 100 at $t=t_*$, when the wind termination 
shock is reached (figures \ref{janwb} and \ref{decwb}), a contrived feature without a physical
foundation. We also find that the wind-bubble scenario requires winds which are as tenuous as those 
for the reverse-forward shock scenario. 

 Thus the reverse-forward shock scenario provides a more natural explanation than the wind-bubble
scenario for the steep early decay of the optical emission of the afterglows 990123 and 021211.
Given that the GRB ejecta can be initially magnetized and that the RS is less relativistic than 
the FS, the microphysical parameters might differ behind the two shocks. If their equality is 
required for a simpler scenario, with fewer assumptions, then a wind-like CBM is favoured by the
reverse-forward shock scenario, though a problem still exists: the low wind density inferred in 
each case ($A_* < 0.1$), which is similar to that derived by Chevalier \etal (2004) for the 
afterglows 020405 and 021211 and by Price \etal (2002) for the afterglow 011211. 
 In the sample of 64 Galactic WR stars analyzed by Nugis \& Lamers (2000) there is only one star 
with $A_* < 0.1$, the majority of the other stars having a mass-loss rate $\dot{M} \in (0.5 - 7) 
\times 10^{-5} \Msunyear$, a wind velocity $v_w \in (1000-3000)\, \kms$, and $A_* \in (0.5,3)$. 
The dependence of the mass-loss rate on stellar mass and metallicity inferred by Nugis \& Lamers 
(2000) lead to $\dot{M} \sim 10^{-6} (M/M_\odot)^{1.1} Y^{2.2}\, \Msunyear$ for WN stars and  
$\dot{M} \sim 10^{-5} (M/M_\odot)^{1.1} Y^2 Z\, \Msunyear$ for WCs, which may suggest that 
the tenuous winds required by the reverse-forward shock scenario for the afterglows 990123 and 
021211 arise WR stars which are less massive and less metal rich than Galactic WRs (Wijers 2001, 
Chevalier \etal 2004). If such stars do not exist, then either the microphysical parameters
must be different behind the RS crossing the GRB ejecta and the FS sweeping-up the CBM or the fast 
declining early optical emission of the afterglows 990123 and 021211 is not arising in the RS. 
One possibility is that the early optical afterglow emission is produced in internal shocks 
occurring in an unsteady wind (\Mesz \& Rees 1999), a scenario which was not investigated in this 
work.

\end{document}